\documentclass[lettersize,journal]{IEEEtran}
\usepackage{cite}
\usepackage{amsmath,amsfonts}
\usepackage{algorithmic}
\usepackage{array}
\usepackage[caption=false,font=normalsize,labelfont=sf,textfont=sf]{subfig}
\usepackage{textcomp}
\usepackage{stfloats}
\usepackage{url}
\usepackage{verbatim}
\usepackage{graphicx,xcolor}
\def\BibTeX{{\rm B\kern-.05em{\sc i\kern-.025em b}\kern-.08em
    T\kern-.1667em\lower.7ex\hbox{E}\kern-.125emX}}
\usepackage{balance}
\begin{document}
\title{Electro-optic frequency combs for multi-wavelength digital holography with high dynamic range}
\author{Leonard Voßgrag, Annelie Schiller, Tobias Seyler, Jens Kießling, Daniel Carl and Ingo Breunig
\thanks{Authors L.V. and I.B. are with University of Freiburg, Department of Microsystems Engineering – IMTEK, Laboratory for Optical Systems, Freiburg, Germany (e-mail: leonard.vossgrag@imtek.uni-freiburg.de);
Authors A.S., T.S., J.K., D.C. and I.B. are with Fraunhofer Institute for Physical Measurement Techniques, IPM, Freiburg, Germany;
Author D.C. is also with University of Freiburg, Institute of Sustainable Systems Engineering – INATECH, Chair Production Control, Freiburg, Germany}}


\maketitle

\begin{abstract}
Multi-wavelength digital holography enables surface-shape measurements with an exceptional dynamic range by combining interferometric resolution with synthetic wavelengths spanning multiple length scales. Although the concept promises measurement ranges of many orders of magnitude, its practical implementation is limited by the lack of light sources that allow fast, reliable, and calibration-free switching between synthetic wavelengths over a wide frequency range.

Here, we present a synthetic-wavelength generator based on an electro-optic frequency comb with electronically tunable modulation frequency and a set of switchable optical band-pass filters. By combining discrete selection of comb-lines with continuous radio-frequency tuning, the proposed scheme merges the advantages of single-sideband modulation and filter-based comb extraction. Using only off-the-shelf fiber-coupled components, the system provides synthetic frequencies from $\approx$~100~MHz to 220~GHz, corresponding to synthetic wavelengths from meters down to millimeters in the visible spectral range. The generator achieves MHz-level frequency accuracy, side-mode suppression exceeding 40~dB, and switching times below 25~ms, even without active stabilization.

We experimentally characterize the spectral purity and frequency agility of the source and demonstrate rapid, programmable tuning of synthetic wavelengths over three orders of magnitude. As a proof of concept, we apply the light source to multi-wavelength digital holography and reconstruct the surface of an industrially machined metal part featuring height variations from 0.1~mm up to 100~mm. The measurements achieve ten-micrometer-level precision using 7 single wavelengths covering synthetic wavelengths from 1.36 mm to 1.874 m within a  acquisition time below 2 seconds.

The presented architecture combines high dynamic height measurement range of 50 dB, fast electronic reconfigurability, and intrinsic frequency calibration, making it a promising light source for high-speed interferometric surface metrology.
\end{abstract}

\section{Introduction}
\IEEEPARstart{T}{he} surface shape of an object can be measured by illuminating it with coherent light at the wavelength $\lambda$ and analyzing the interference pattern formed by the superposition of the wave reflected from the surface and a reference wave. In its conventional form, interferometry provides outstanding resolution, typically on the order of $\lambda/100$. However, the unambiguous range is limited to $\lambda/2$. As a result, the dynamic range of a single-wavelength interferometric measurement spans only about two decades - hardly impressive when compared with that of an ordinary ruler.

Multi-wavelength interferometry (MWI) offers a route to overcome this limitation by introducing synthetic wavelengths \cite{Hildebrand67}
\begin{equation}
\Lambda=\frac{c}{|\nu_2-\nu_1|}\equiv\frac{c}{F}\;,
\label{eq:1}
\end{equation}
derived from the speed of light $c$ and the frequency difference $F=|\nu_2-\nu_1|$ of two coherent optical fields. Here, $F$ can be interpreted as a synthetic beat-frequency.

By performing measurements at several synthetic wavelengths and combining the resulting phase maps through hierarchical phase unwrapping, multi-wavelength interferometry achieves a dynamic measurement range spanning up to ten orders of magnitude — from micrometers to hundreds of meters — while maintaining sub-micrometer resolution \cite{Wagner00}. This extraordinary range is realized within a single physical principle, establishing multi-wavelength interferometry as a uniquely powerful tool for surface metrology.

Despite this tremendous potential, the practical implementation of MWI remains limited by the lack of suitable light sources that can reliably switch between synthetic wavelengths across several orders of magnitude \cite{Osten22,Yang18}. Two main strategies are commonly used to realize light sources for multi-wavelength interferometry: (i) combining several free-running lasers operating at different frequencies and (ii) tuning a single laser over a defined frequency range.

The first approach relies on multiple independent lasers whose optical frequencies or frequency differences must be accurately known. This restricts the practically achievable synthetic wavelengths to the millimeter scale or below, although the concept has matured into robust industrial systems \cite{Fratz21}.

The second approach offers greater flexibility by continuously tuning a single laser, yet reliable operation demands active frequency tracking and feedback control. Moreover, tuning speeds remain rather slow: for instance, a tunable continuous-wave parametric oscillator achieved synthetic wavelengths from 2~\textmu m to 2~m - a remarkable range \cite{Seyler22} — but switching between two target wavelengths requires several minutes.

Thus, both concepts are constrained either by calibration effort and/or by tuning speed. Neither enables fast, electronically controlled switching between synthetic wavelengths across multiple orders of magnitude — a prerequisite for fully exploiting the potential of multi-wavelength interferometry.

Electro-optically generated frequency combs can be used to overcome this unsatisfactory limitation. They produce a set of phase-coherent optical sidebands around a single laser frequency $\nu_0$, with a defined frequency spacing determined by the radio-frequency (RF) drive $f$. The optical frequencies within the comb are given by
$\nu_N=\nu_0+N f$, where $N$ is an integer denoting the comb-line order. The frequency spacing - and hence the calibration of the comb - is defined entirely by the frequency stability of the RF source. As a result, no wavemeter or optical reference is required. To employ such frequency combs in multi-wavelength interferometry, the optical spectrum must be converted into tunable single-frequency light, such that the resulting frequency differences — and hence the synthetic wavelengths — are determined solely by $N$ and $f$, i.e. $F=c/\Lambda=|N_2f_2-N_1f_1|$.

\begin{figure*}[!t]
\centering
\includegraphics[width=182mm]{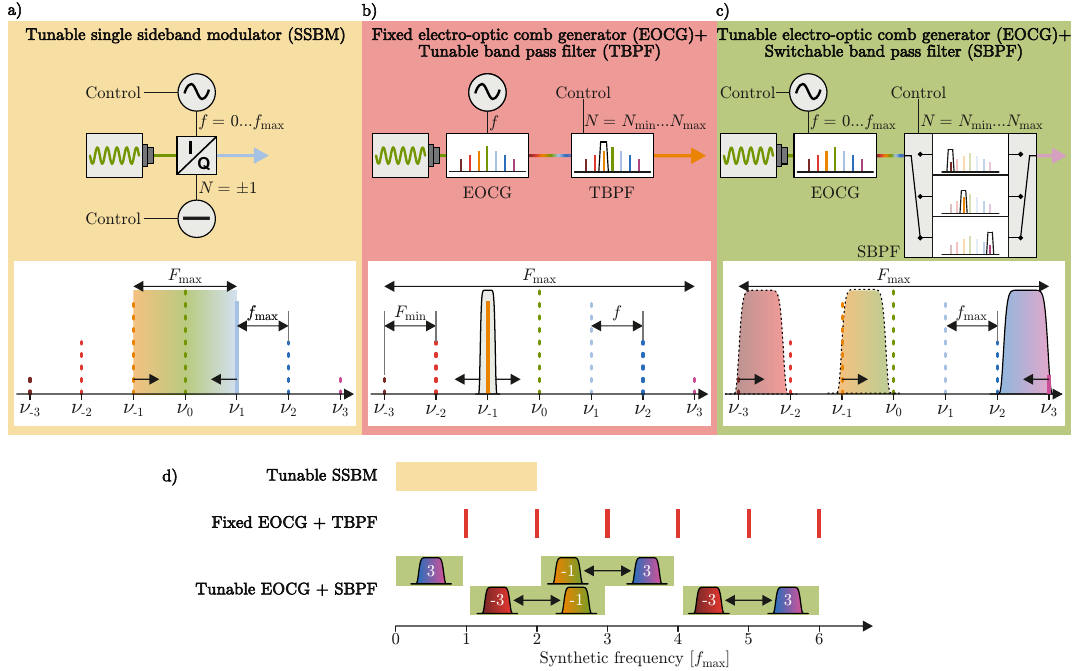}
\caption{a)-c): Different schemes for modulated light sources with their available control parameters of the modulation frequency $f$ and or the sideband order N. EOCG, electro-optic comb-generator; TBPF, tunable bandpass filter; BPF, bandpass filter. d): Overview of the available synthetic wavelength based on the presented schemes.}
\label{fig1}
\end{figure*}

Recently, we have demonstrated that a single-sideband (SSB) generator can be used to achieve this conversion \cite{Vossgrag25} (see Fig.~\ref{fig1}a). In this approach, four electro-optic comb generators are interferometrically coupled such that the output spectrum contains only a single dominant component at $N=\pm1$. Its optical frequency can be tuned continuously between $\nu_0-f_\mathrm{max}$ and $\nu_0+f_\mathrm{max}$, resulting in a maximum synthetic frequency of $F_\mathrm{max}=2f_\mathrm{max}$. The minimum accessible synthetic frequency $F_\mathrm{min}$ is determined by the uncertainty $\Delta\nu$ of the carrier frequency $\nu_0$ during the measurement time. Accepting a relative uncertainty of 1~\%, we obtain $F_\mathrm{min}\approx100\Delta\nu$. In a proof-of-concept experiment, the modulator was operated up to $f_\mathrm{max}=10$~GHz with a carrier-frequency uncertainty of $\Delta\nu\approx1$~MHz \cite{Vossgrag25}. This corresponds to a synthetic-frequency range $F_\mathrm{max}/F_\mathrm{min}\approx200$ and consequently, to a dynamic height-measurement range of about $10^4$.

Access to higher synthetic frequencies can be achieved by combining electro-optic comb generators, operated at a fixed RF frequency $f$ with tunable optical filters \cite{Hase21,Hase23} (see Fig.~\ref{fig1}b). These filters extract a single comb component at a target order between $N=-N_\mathrm{max}$ and $N=+N_\mathrm{max}$. The corresponding maximum synthetic frequency is $F_\mathrm{max}=2N_\mathrm{max}f$, while the minimum value is $F_\mathrm{min}=f$. Hence, the dynamic range in synthetic frequency equals the number $2N_\mathrm{max}$ of accessible sidebands and is independent of $f$. In proof-of-concept experiments, radio frequencies around $f=10$~GHz and up to $2N_\mathrm{max}\approx300$ accessible sidebands were reported \cite{Hase21, Hase23}, resulting in synthetic frequencies reaching the terahertz range. The resulting dynamic range in synthetic frequency  however, is of the same order as that obtained with a single-sideband modulator operated at the same RF frequency.

Despite their striking advantage of intrinsic wavelength calibration — relying solely on the stability of the RF reference — the current state-of-the-art electro-optic frequency-comb-based synthetic wavelength generators still fall short of the extraordinary dynamic range that multi-wavelength interferometry offers.

Here, we demonstrate a synthetic-wavelength generator based on an electro-optic modulator driven by a variable radio frequency and combined with several fixed band-pass filters (see Fig.~\ref{fig1}c). Each filter extracts a single comb component $\nu_N$ at its respective target order $N$. By varying the modulation frequency $f$, the optical frequency $\nu_N=\nu_0+N f$ can be tuned continuously across the passband of the selected filter. Two optical switches are used to route the output to the desired order $N$. In the example depicted, the system can switch between $\nu_{-3}$, $\nu_{-1}$ and $\nu_{+3}$. Combining the discrete selection of $N$ with the continuous tuning of $f$ allows the output frequency to be set to almost any value between $\nu_0+N_\mathrm{min}f_\mathrm{max}$ and $\nu_0+N_\mathrm{max}f_\mathrm{max}$ (see Fig.~\ref{fig1}d). Consequently, the maximum achievable synthetic frequency is $F_\mathrm{max}=2N_\mathrm{max}f_\mathrm{max}$, i.e. the same scaling as in filter-based EO-comb generators. At the same time, the continuous tunability provided by the variable RF drive reduces the minimum synthetic frequency to $F_\mathrm{min}\approx100\Delta\nu$ determined by the carrier-frequency uncertainty during the measurement time — identical to the SSB-based approach. 
\begin{figure*}[!t]
\centering
\includegraphics[width=182mm]{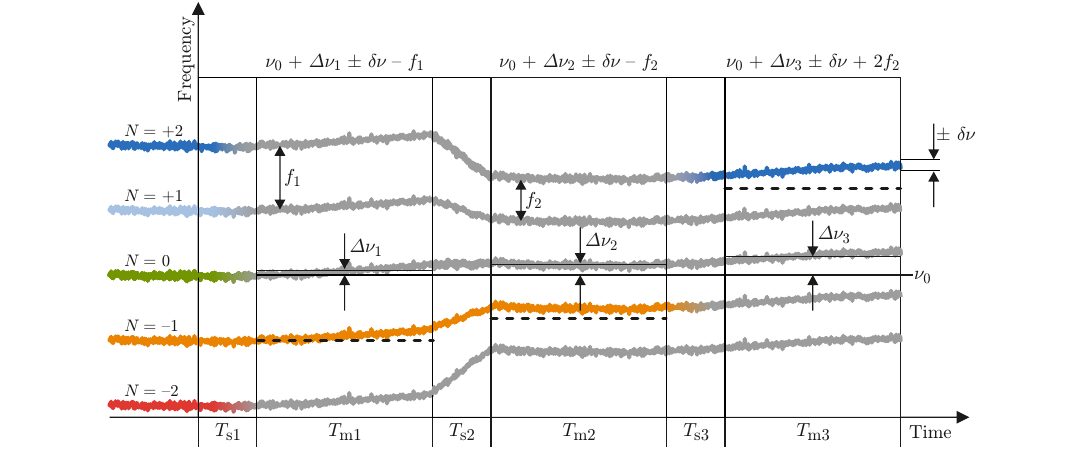}
\caption{Temporally varying frequencies of the components of an electro-optic frequency comb pumped at the optical carrier frequency $\nu_0$ and modulated at the radio-frequency values $f_{1,2}$, in combination with two switchable band-pass filters. Shown are three consecutive acquisition cycles consisting of alternating switching and measurement intervals where every but the selected component are sketched in gray. During the switching times $T_\mathrm{s1,2,3}$, either the modulation frequency $f$ or the selected filter is changed to address a different comb component at the order $N$. The intervals $T_\mathrm{m1,2,3}$ denote the time windows during which interferograms are recorded while a nominal comb component is selected. Due to the laser-frequency drift $\Delta\nu$ and the short-term frequency fluctuations $\delta\nu$ ( both not to scale), the actual optical frequency during the measurement intervals can significantly differ from the target values (dashed lines).}
\label{fig2}
\end{figure*}

Our scheme therefore combines the strengths of both prior architectures: it adopts the large maximum synthetic frequency of fixed-frequency, filter-based comb systems while retaining the small minimum synthetic frequency of variable-frequency SSB generators. As a result, the dynamic measurement range is increased by a factor corresponding to the number of accessible filter orders. Using the parameters discussed above, this translates to a dynamic measurement range of approximately $10^6$, enabling the measurement of meter-scale surface variations with micrometer-level uncertainty. A comparable approach has been successfully employed by  Li et al. \cite{Li25} for one-dimensional ranging. We expand this approach to be usable for holographic surface metrology.

In this proof-of-concept study, we experimentally demonstrate a system based on this approach using only off-the-shelf components. We show that it reliably generates synthetic wavelengths from the millimeter to the meter scale, i.e. across three orders of magnitude, and that it is fully suitable for multi-wavelength interferometry. As a demonstration, we reconstruct the surface shape of a milled metal part featuring structures ranging from $100$~\textmu m to  100~mm.

\section{Requirements on laser stability and spectral filtering}
\noindent The approach presented in this work relies on combining electro-optic frequency comb generation with selectable spectral filtering to produce tunable single-frequency light for multi-wavelength interferometry. Its reliable operation requires that two key conditions are fulfilled. First, the frequency drift of the underlying laser must remain sufficiently small during the acquisition of successive interferograms. Second, the employed optical filters must provide adequate suppression of unwanted sidebands to ensure spectral purity. In the following, we discuss these requirements and derive practical criteria for their fulfillment.
\subsection{Frequency Drift}
Figure~\ref{fig2} illustrates the temporal evolution of the optical frequencies generated by an electro-optic frequency comb during three consecutive acquisition cycles, targeting the synthetic frequencies $F_1=f_1$, $F_2=f_2$, $F_3=2f_2$. Throughout the entire sequence, the pump laser frequency $\nu_0$ exhibits a slow drift $\Delta\nu$ and short-term fluctuations $\delta\nu$. In the first switching interval $T_\mathrm{s1}$, all comb components are suppressed except the one with order $N=-1$. During the subsequent measurement interval $T_\mathrm{m1}$, interferograms are recorded at an actual optical frequency $\nu_1=\nu_0+\Delta\nu_1\pm\delta\nu-f_1$. In the second switching interval, the modulation frequency is changed from $f_1$ to $f_2$. Consequently, during the following measurement interval $T_\mathrm{m2}$, the optical frequency becomes $\nu_2=\nu_0+\Delta\nu_2\pm\delta\nu-f_2$. For the third cycle, the band-pass filter is switched to transmit the component with order $N=2$. During the measurement interval $T_\mathrm{3}$, the optical frequency is therefore $\nu_3=\nu_0+\Delta\nu_3\pm\delta\nu+2f_2$. The target synthetic frequencies $F_1=\nu_0-\nu_1$, $F_2=\nu_0-\nu_2$ and $F_3=\nu_3-\nu_0$ are obtained only if the drift $\Delta\nu$ and the short-term fluctuations $\delta\nu$ remain small compared with the target synthetic frequency $F$.

For a typical free-running single-frequency laser, short-term fluctuations are on the order of 1~MHz, while long-term drift can reach 100~MHz/h. If the total acquisition time is limited to a few seconds, the uncertainty of the synthetic frequency is therefore dominated by the short-term fluctuations, i.e. by the laser linewidth. Assuming a linewidth of 1~MHz and a target relative uncertainty below 1~\%, the minimum practical synthetic frequency is on the order of 100~MHz, corresponding to synthetic wavelengths in the meter range. For frequency-stabilized lasers, substantially lower synthetic frequencies, and thus larger synthetic wavelengths, become accessible.

\begin{figure}[!t]
\centering
\includegraphics[width=90mm]{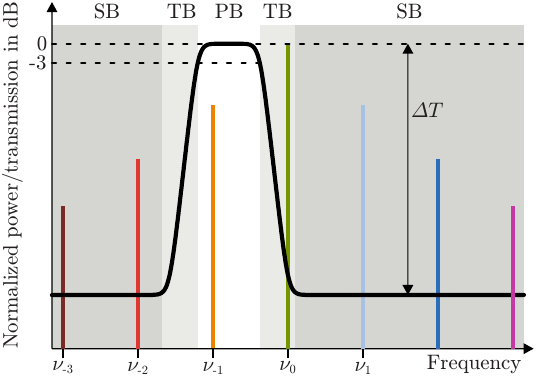}
\caption{Normalized power of selected frequency-comb components at the optical frequencies $\nu_{-3\ldots 3}$ and normalized transmission of the optical band-pass filter as a function of frequency. PB denotes the 3 dB passband of the filter. SB indicates the stopband regions, characterized by the stopband suppression $\Delta T$, while TB marks the transition bands between passband and stopband.}
\label{fig:Filter}
\end{figure} 
\subsection{Suppression of unwanted sidebands}
In the following, we address the suppression of unwanted side modes by optical filtering. An electro-optic frequency-comb generator produces a set of sidebands $\nu_N$ centered around the optical carrier frequency $\nu_0$ and separated by the RF modulation frequency $f$. The relative power of the individual comb components depends on both the RF driving power and the comb order $N$.

To maintain maximum flexibility in the selection of optical frequencies, the spectral width of the filter passband (PB) should be comparable to the maximum RF frequency $f_\mathrm{max}$. However, in any practical optical filter, the passband and stopband (SB) are separated by a finite transition band (TB), in which the transmission decreases gradually rather than abruptly (see Fig.~\ref{fig:Filter}).

To illustrate the resulting constraints, we consider a realistic scenario. The maximum RF frequency is $f_\mathrm{max}=20$~GHz, and the power of the unwanted side mode exceeds that of the target component by a factor of ten. In a previous study employing the same interferometer and reconstruction algorithms, we demonstrated that unwanted side modes of an electro-optically driven synthetic-wavelength generator must be suppressed by at least 20~dB for reliable application in multi-wavelength interferometry \cite{Vossgrag25}. Consequently, the optical filter must provide a total stopband suppression of $\Delta T=30$~dB.

Typical fiber-based optical band-pass filters providing such suppression levels exhibit transmission slopes of several dB per gigahertz. As a result, the transition band spans approximately 10~GHz. The width of this transition band is therefore comparable to the maximum RF frequency and cannot be neglected when using commercially available filters.

It is important to note that the same conclusion holds even if the electro-optic frequency comb exhibits equal power in all components. In this case, the required stopband suppression is reduced from 30~dB to 20~dB, which corresponds to only a modest reduction of the transition-band width by a factor of approximately 1.5. Thus, the transition band remains on the order of 10~GHz.

When several filters are stacked in series, their transmission functions multiply, leading to a substantially increased out-of-band suppression and an effectively steeper roll-off. While the passband is moderately reduced, the improvement in side-mode suppression within the transition region is significantly stronger. This approach therefore enables the required suppression of unwanted comb components without relying on unrealistically sharp single-filter characteristics.

These considerations discussed in this section define the practical conditions under which the proposed synthetic-wavelength generator can be operated reliably.
\section{Generation of synthetic wavelengths}
In this section, we describe the experimental realization of the synthetic-wavelength generator introduced above. We first present the experimental setup and subsequently characterize its spectral purity and frequency tuning behavior.
\subsection{Experimental Setup}
\begin{figure*}[!h]
\centering
\includegraphics[width=182mm]{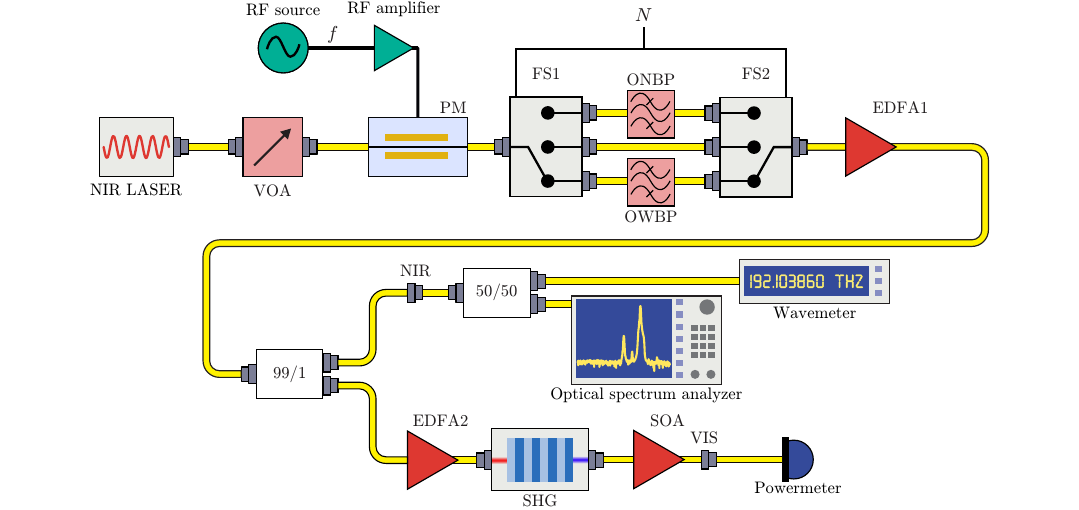}
\caption{Experimental setup for generating synthetic wavelengths and its characterization: NIR, near-infrared; VOA, voltage-controlled optical attenuator; RF radio frequency; PM, phase modulator; FS, fiber switch; ONBP, optical narrow bandpass filter; OWBP, optical wide bandpass filter; EDFA, erbium doped fiber amplifier; SHG second harmonic generation; SOA semiconductor optical amplifier; VIS, visible light.}
\label{fig:Setup1}
\end{figure*}

\noindent Fig.~\ref{fig:Setup1} shows a sketch of the experimental setup comprising the synthetic-wavelength generator as well as additional instruments used for its characterization.
Near-infrared light at the optical frequency $\nu_0=192.144$~THz, emitted by a free-running fiber laser (Koheras, BasiK E15), is used as the optical carrier. The laser exhibits a linewidth of approximately 100~kHz. Its output power is adjusted between 0 and 15~mW using a voltage-controlled optical attenuator (Thorlabs, VOA1000) and serves as the input to an electro-optic phase-modulator-based frequency-comb generator (Exail, MPZ-LN-40). The radio-frequency (RF) drive signal in the range $f=30\ldots40$~GHz is generated by a microwave signal generator (Anritsu, MG3696A) and amplified by an RF amplifier (Exail, DR-AN-40-MO) before being applied to the phase modulator. The modulated optical signal is subsequently routed to a switchable filter bank providing three selectable output configurations.

In the first configuration, a narrowband optical filter with a 4.5 GHz passband centered at 192.214~THz (Exail, IXC-FBG-PS-M) transmits the comb component at order $N=2$. In the second configuration, two cascaded flat-top ITU-grid dense wavelength-division multiplexing (DWDM) filters aligned to ITU channel 21 (192.100 THz frequency) with a passband width of 10~GHz (Lightel, 50~GHz DWDM Band Pass Filter) are used to transmit the component at order $N=-1$. In the third configuration, the optical carrier ($N=0$) is accessed by setting the RF drive to zero, without employing any optical filter. Switching between these configurations is realized using two MEMS-actuated fiber-optic switches (Fibermart, PMOSW-1x4M-X).

The optical power at the output of the filter bank is equalized using an erbium-doped fiber amplifier (Thorlabs, EDFA300P) to compensate for the non-uniform power distribution of the comb components as well as insertion losses introduced by filters and switches. A -20 dB fiber tap coupler extracts 1~\% of the optical power for characterization purposes, allowing the spectral purity to be monitored with an optical spectrum analyzer (Yokogawa, AD6370) and the absolute optical frequency to be measured using a wavemeter (HighFinesse, WS7-60). The remaining optical power is amplified by a second EDFA (IPG Photonics, EAD-3-C-PM).

The amplified near-infrared light is subsequently frequency-doubled in a nonlinear crystal module (HC Photonics, PMC2307050016). The generated visible radiation around a wavelength of 780~nm is further amplified using a semiconductor optical amplifier (Thorlabs, BOA780P), which also compensates for the frequency-dependent conversion efficiency of the frequency-doubling stage.

The setup comprises several optical amplification stages, whose roles are clarified by the following typical power levels. Approximately 10~mW of optical power is launched into the phase modulator. At the output of the filter bank, the optical power is about 10 mW when transmitting the carrier ($N=0$), approximately 0.2~mW for the first negative sideband ($N=-1$), and about 0.01~mW for the second positive sideband ($N=2$). By adjusting the pump power of the first EDFA, the power of all selected components is equalized to approximately 10~mW. The second EDFA increases this level to about 200~mW at the input of the frequency-doubling module, yielding 1 to 2~mW of visible output power depending on the optical frequency. The final amplifier equalizes the visible output power to approximately 5~mW.

At first glance, using multiple optical amplification stages and frequency doubling increases the system's complexity. However, this architecture provides several key advantages. First, frequency doubling effectively doubles the bandwidth of the generated frequency comb without requiring additional high-frequency RF hardware. Second, since the efficiency of second-harmonic generation scales with the square of the input power, this stage enhances the spectral purity of the generated light, as demonstrated by Sabatti et al. \cite{Sabatti24}. Finally, the use of frequency doubling enables the combination of mature, off-the-shelf telecommunication components in the near-infrared with cost-effective silicon-based cameras in the visible spectral range. These cameras offer small pixel sizes and large sensor areas, thereby improving the achievable lateral resolution compared with near-infrared camera arrays based on gallium arsenide.

\subsection{Spectral purity and frequency-switching performance}
\noindent We now turn to the experimental characterization of the synthetic-wavelength generator, focusing on its spectral purity and frequency-tuning behavior, which constitute the key performance metrics for applications in multi-wavelength interferometry, as motivated above.

The suitability of a frequency-comb-based synthetic-wavelength generator is primarily determined by its spectral purity. As discussed earlier, multi-wavelength interferometry requires single-frequency operation. We therefore investigate the residual presence of unwanted frequency components after the extraction of the desired comb lines.
\begin{figure*}[!t]
\centering
\includegraphics[width=182mm]{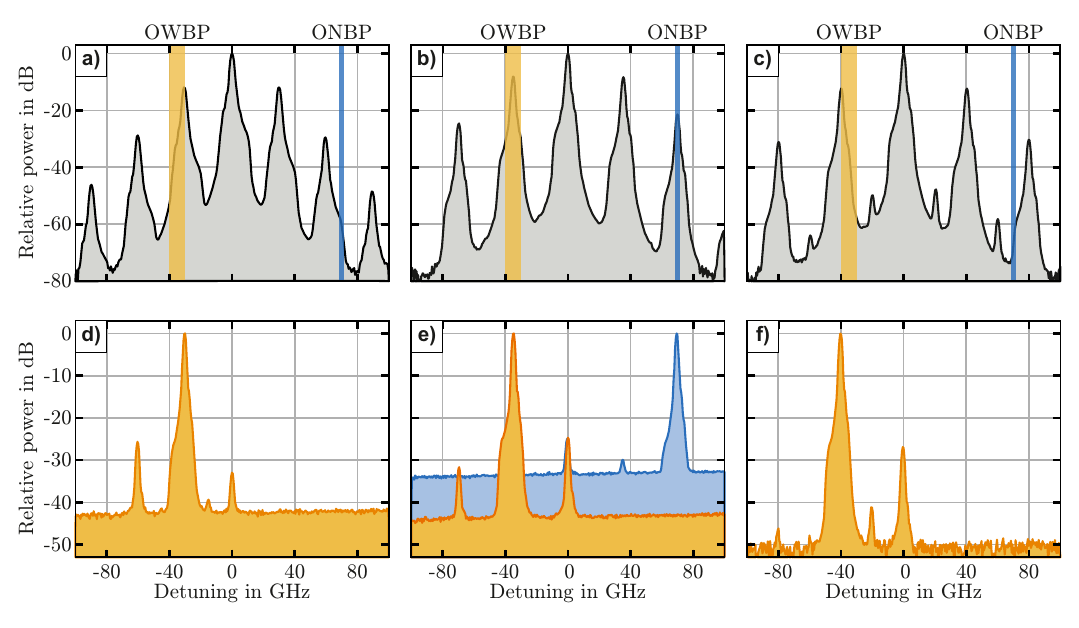}
\caption{Near-infrared optical spectra of the frequency comb (gray traces in a, b and c) and of the light at the output of the filter bank (yellow and blue traces in d, e and f). The yellow traces indicate the output at $N=-1$ transmitted by the optical wide-band filter (OWBF), while the blue one indicates the output at $N=2$ transmitted by the optical narrow-band filter (ONBF). The modulation frequency is set to 30~GHz (a and d), to 35~GHz (b and e) and to 40 GHz (c and f).}
\label{fig4}
\end{figure*}
Figure~\ref{fig4} shows optical spectra recorded at modulation frequencies of $f=30$~GHz (a, d), $f=35$~GHz (b, e) and $f=40$~GHz (c, f). Subfigures (a-c) display the spectra at the output of the electro-optic phase modulator, while subfigures (d-f) show the corresponding spectra after transmission through the optical filter stages.

The wide-band filter transmits the negative first-order sideband over the entire range $f=30~\ldots~40$~GHz, while suppressing unwanted side modes to levels below -20~dB relative to the target component. In contrast, the narrow-band filter transmits the positive second-order component at $f=35$~GHz, achieving a suppression of unwanted side modes exceeding -30~dB. For modulation frequencies of $f=30$~GHz and $f=40$~GHz, no significant optical power is transmitted through the narrow-band filter, consistent with its limited passband.

We do not observe any unwanted sidebands if we switch off the RF power and setting the filter bank to transmit the carrier frequency.  These observations confirm that the required side-mode suppression is met for the relevant operating points, enabling single-frequency operation at $N=-1$, $N=0$ and $N=2$ within the accessible RF range. Furthermore, one has to take into account that subsequent second-harmonic generation will increase the suppression of unwanted sidebands to more than 40~dB for the light in the visible range around 780 nm wavelength \cite{Sabatti24}.

Having established single-frequency operation at the relevant output states, we quantify the frequency agility of the system, i.e. how fast and how accurately the selected output frequency can be switched. While the output frequencies corresponding to $N=0$ and $N=2$ are fixed at $\nu_0$ and $\nu_0+70$~GHz, respectively, the frequency of the $N=-1$ component can be tuned continuously over a range of 10~GHz by varying the modulation frequency. Ideally, when considering only this component, the temporal evolution of the synthetic frequency $F(t)$ can follow an arbitrary waveform within this tuning range.\\
To experimentally characterize this capability, we map an acoustic signal $f_\mathrm{note}(t)$ onto a target synthetic frequency in the near infrared according to
\begin{equation}
F(t)=10^{f_\mathrm{note}(t)/261.63~\mathrm{Hz}}\times 1~\mathrm{MHz},
\label{eq:BR}
\end{equation}
where the exponential mapping ensures a wide dynamic range of synthetic frequencies.

\begin{figure*}[!t]
\centering
\includegraphics[width=182mm]{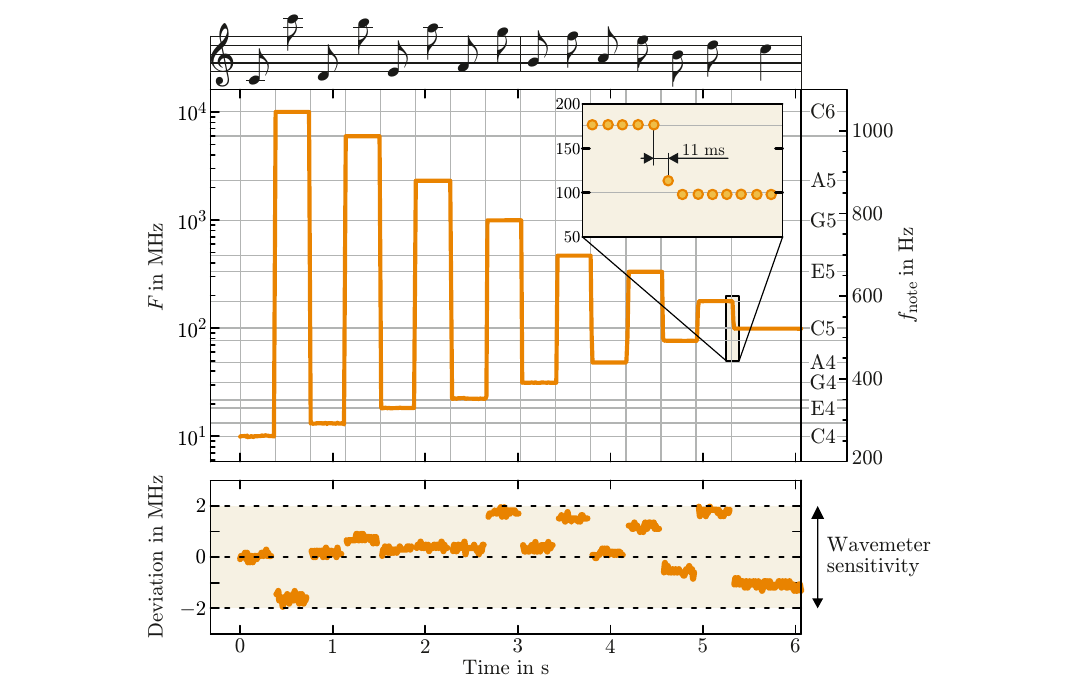}
\caption {Temporal mapping of an acoustic signal onto a synthetic-frequency trajectory.
The time-dependent acoustic signal $f_\mathrm{note}(t)$, corresponding to two bars of interleaved C-major scales, is converted into a target sequence of synthetic frequencies $F(t)$
according to Eq.~(\ref{eq:BR}) (gray line). Orange markers denote the experimentally measured values. The lower panel shows the point-by-point deviation between the measured and target frequencies.}
\label{fig:BR}
\end{figure*}

As the acoustic test signal, we arranged two bars of two interleaved c-major scales. One scale plays up from C4 to C5, the other one is descending from C6 to C5.  At a tempo of 80 beats per minute the resulting duration of the total sequence is 6~s. The lowest note in this sequence is C4 at $f_\mathrm{note}=261.63$~Hz, corresponding to a minimum synthetic frequency $F_\mathrm{min}=10$~MHz. The highest note is C6 at $f_\mathrm{note}=1046.52$~Hz, yielding a maximum synthetic frequency of $F_\mathrm{max}=10$~GHz. This corresponds to a dynamic range of $F_\mathrm{max}/F_\mathrm{min}=1000$, considering only the output at $N=-1$.

Experimentally, the desired temporal variation $F(t)$ is realized by setting the modulation frequency to $f(t)=30~\text{GHz}+F(t)$. During this sequence, the actual optical output frequency of the $N=-1$ component is continuously monitored using a wavemeter with a temporal resolution of 11~ms and compared with the target behavior defined by Eq.~(\ref{eq:BR}). The resulting time traces are shown in Fig.~\ref{fig:BR}.

The experimentally measured output frequency nicely reproduces the prescribed target sequence 
$F(t)$ throughout the entire 6~s interval, covering synthetic frequencies from 10~MHz up to 10~GHz. The lower panel of Fig.~\ref{fig:BR} shows the point-by-point deviation between  measured and target frequencies. Over the full sequence, the residual deviations remain below 2~MHz, without any systematic drift or accumulation of error. This confirms that the tuning accuracy is not limited by the modulation or switching scheme, nor by any drift of the laser frequency, but rather by the resolution of the wavemeter used for readout.

Rapid frequency changes, corresponding to transitions between notes in the acoustic signal, are  reproduced without observable overshoot or ringing beyond 2 consecutive samples in a time of 22~ms. This value is as expected, as we approaching the limit of our RF-sources switching time ($\leq$ 25~ms). This value represents the maximum switching-time of the whole setup, as the fiber switches are specified $\leq$10~ms for this parameter.   

Taken together, these results demonstrate that the synthetic-wavelength generator provides fast (below 25~ms), accurate (MHz-level), and repeatable frequency tuning in the near infrared over nearly three orders of magnitude using a single output state ($N=-1$), thereby fully meeting the requirements derived in Sec.~II.

Combining these results with the two additional output states, the following synthetic frequencies are experimentally accessible in the near-infrared. Using the tunable $N=-1$ component alone, synthetic frequencies in the range $0.010\ldots10$~GHz are obtained. By combining the carrier ($N=0$) with the $N=-1$ component, synthetic frequencies of $30\ldots40$~GHz are accessible. A fixed synthetic frequency of $70$~GHz is obtained by combining the carrier with the $N=2$ component, while synthetic frequencies of $100\ldots110$~GHz result from combining the $N=-1$ and $N=2$ components.

Taken together, these operating points provide a synthetic-frequency dynamic range exceeding three orders of magnitude, even when allowing for a relative uncertainty margin of 1~\%. Assuming a frequency uncertainty at the MHz level, this corresponds to a practical range of 
$F\approx0.1\ldots$110~GHz. Subsequent frequency doubling extends this range by a factor of two. In the visible spectral range, the system therefore covers synthetic wavelengths from approximately 1.4~mm to 1.5~m, while maintaining side-mode suppression exceeding 40~dB and an output power of about 5~mW.

It is important to note that the experimental setup presented here does not exploit the fundamental performance limits of the proposed scheme. With respect to the tuning range, the maximum achievable synthetic frequency—and thus the minimum synthetic wavelength—can be significantly increased by raising the RF modulation frequency $f$ and/or the applied RF power. Increasing $f$ enlarges the spectral separation between adjacent comb components, while higher RF power enables the use of sidebands with larger comb orders $N$. Li et al. have demonstrated that electro-optic phase modulators can provide sufficient optical power even for sidebands with orders up to $N=\pm10$\cite{Li25}. Combining such a frequency comb with a modulation frequency of $f=50$~GHz would allow near-infrared synthetic frequencies approaching 1~THz and, consequently, synthetic wavelengths down to approximately 150~\textmu m in the visible spectral range. In particular, electro-optic modulators based on thin-film lithium niobate \cite{Wang18} are well suited for this purpose, as they can be operated at RF frequencies exceeding 100~GHz while requiring substantially lower RF drive powers than those used in this proof-of-concept study.

\begin{figure*}[!b]
\centering
\includegraphics[width=182mm]{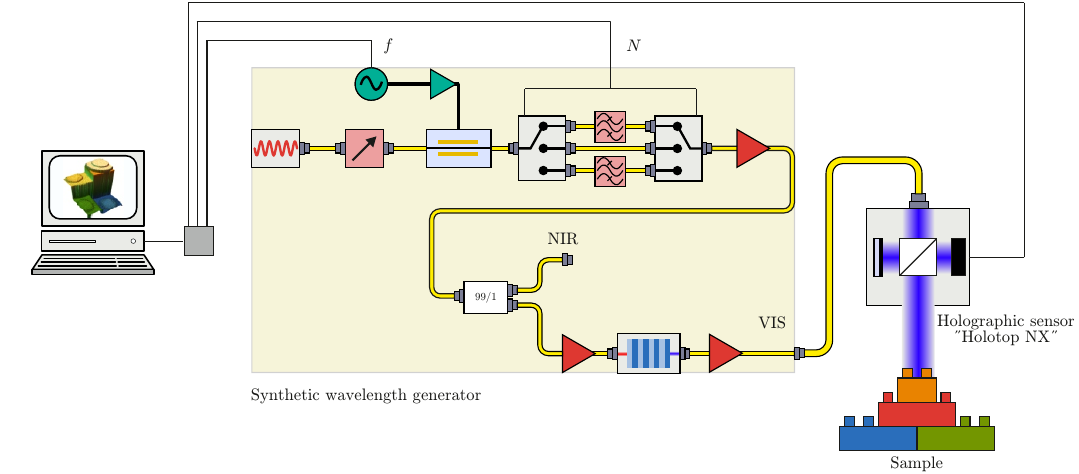}
\caption{Sketch of the measurement setup for 3D surface determination: Our synthetic wavelength generator is connected to the interferometric sensor head, the sample is placed underneath. A computer handles the reconstruction of the holograms as well as the measuring sequence. }
\label{fig5}
\end{figure*}

The relative coverage of synthetic frequencies between $F_\mathrm{min}$ and $F_\mathrm{max}$
can be further increased by incorporating additional optical wideband filters. Operating at modulation frequencies of 50 or 100~GHz is especially attractive in this context, as it enables the direct use of readily available fiber-based ITU-grid band-pass filters.

The achievable switching speed is determined by the performance of both the RF signal generator and the optical switching elements. Modern RF signal generators can provide frequency switching times well below 100~\textmu s \cite{Banerjee20,Dutta24}. Replacing MEMS-based fiber switches with semiconductor optical amplifier based modules would allow switching between different filter paths on sub-microsecond timescales \cite{Tanaka09}. Consequently, overall switching times on the order of 100~\textmu s are achievable using standard off-the-shelf components.

Significant improvements are also possible with respect to the available optical power. The frequency-doubling stage employed in our setup can be operated at input powers of several watts. Replacing the 200~mW amplifier used here with a watt-level optical amplifier would therefore increase the conversion efficiency by approximately an order of magnitude. As a result, visible output powers in the range of several hundred milliwatts at a wavelength of 780~nm could be achieved even without an additional semiconductor optical amplifier.

Taken together, an optimized implementation of the proposed scheme would enable reliable and flexible tuning of synthetic wavelengths from the sub-millimeter regime to meter scales, corresponding to a synthetic wavelength dynamic range of approximately $10^4$, while providing optical output powers on the order of 100~mW. This, in turn, would allow micrometer-scale deformation measurements on meter-sized objects, i.e. a total measurement dynamic range approaching $10^6$, within acquisition times of seconds or less.

\section{Multi wavelength digital holography with high dynamic range}
\noindent Having established the experimental realization, performance, and scalability of the synthetic-wavelength generator, we now turn to its application in multi-wavelength digital holography. In the following section, we demonstrate how the available synthetic wavelengths are employed for hierarchical phase unwrapping and surface-shape measurements across multiple length scales.

For this purpose, we connect our synthetic-wavelength generator to a commercially available interferometric sensor head (Fraunhofer IPM, Holotop NX NIR) as it can be seen in Fig.\ref{fig5}. The holographic measurement setup is essentially a Mach-Zehnder type interferometer, where the reference mirror is mounted on an piezoelectric actuator to retrieve the phase of the hologram via temporal 3-step phase-shifting\cite{Greivenkamp84,Cai04}. This means, three interferograms are acquired at each output frequency of the light source. The whole measuring process is implemented by the associated software to the sensor head (Fraunhofer IPM, HoloSoftware) on a personal computer (PC). Besides the phase retrieval, the software controls the hierarchical unwrapping  and numerical propagation of the acquired data. Reference~\cite{Carl09,Stevanovic21} give further details about the used sensor head and algorithms used.         

\begin{figure*}[!b]
\centering
\includegraphics[width=182mm]{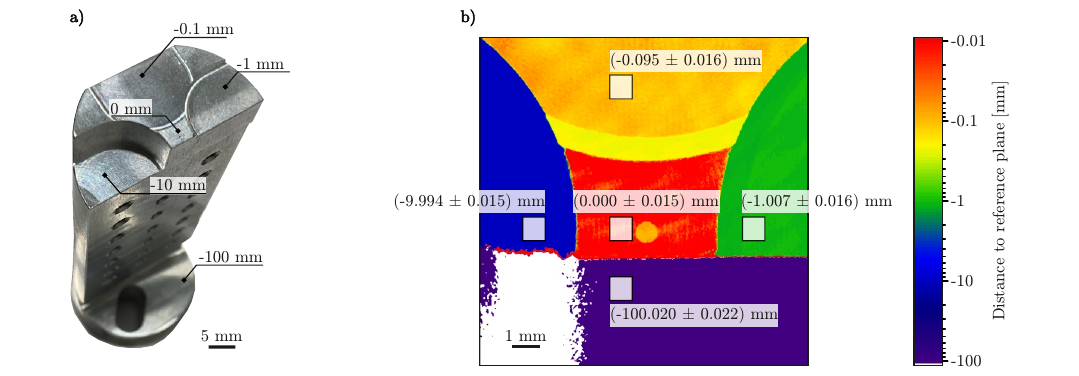}
\caption{a): Photograph of the investigated industrially machined sample made of aluminum,  comprising height features between 0.1 and 100~mm. b): Measured height-map after the use of 5 cascaded synthetic wavelength. The measured values for each plane are state as mean and standard deviation of 200~$\times$~200 pixel regions of interest, indicated by the respective rectangles.}
\label{fig6}
\end{figure*}

To reveal the high dynamic range of the proposed approach, we investigate the surface of one industrially machined part made of aluminum with logarithmic height-features ranging from 0.1 to 100~mm in factors of 10, as it is displayed in Fig.~\ref{fig6}a. These nominal values are confirmed by a tactile measurement with a maximum deviation of 6~µm (Hexagon, Leitz Infinity).  

\begin{table}[!t]
\caption{ Exact synthetic wavelengths $\Lambda_\mathrm{vis}$ from the visible light used for multi-wavelength interferometry and corresponding synthetic frequencies $F_\mathrm{vis, nir}$ from the visible and the near-infrared light along with the respective modulation frequencies $f_{1,2}$ and component orders $N_{1,2}$. \label{tab:Synthetic}}
\centering
\begin{tabular}{|r|r|r|r|r|r|r|}
\hline
$\Lambda_\mathrm{vis}$/mm & $F_\mathrm{vis}$/GHz&$F_\mathrm{nir}$/GHz&$f_1$/GHz&$N_1$&$f_2$/GHz&$N_2$\\
\hline
1873.70 & 0.16 &0.08&40.00&-1&39.92&-1\\
374.74 & 0.80 &0.40&40.00&-1&39.60&-1\\
74.95 & 4.00 & 2.00&40.00&-1&38.00&-1\\
14.99&20.00&10.00&40.00&-1&30.00&-1\\
3.75&80.00&40.00&40.00&-1&0.00&0\\
1.36&220.00&110.00&40.00&-1&35.00&+2\\
\hline
\end{tabular}
\end{table}

For this investigation, we use 6 synthetic wavelengths $\Lambda_\mathrm{vis}$ between 1.36 and roughly 1874~mm from the visible light generated by the light source described above. The exact values of the synthetic wavelength, corresponding synthetic frequencies $F_\mathrm{vis}$ from the visible light as well as the synthetic frequencies $F_\mathrm{nir}$ from the near infrared light along with the respective modulation frequencies $f_{1,2}$ and component orders $N_{1,2}$ are given in Tab.~\ref{tab:Synthetic}. It can be seen that the six synthetic wavelengths are formed by a total number of seven optical light fields. In particular, $f_1=40$~GHz and $N_1=-1$ are constant, while the values for $f_2$ and $N_2$ are varied. For each of the seven optical fields, three interferograms are recorded and subsequently reconstructed to a height map.

The overall time including changing of the optical frequencies and capturing the interferograms was 1.5~s. The accumulated switching time of the light-source accounts for 0.15~s of the total acquisition-time. The cameras exposure time is 12~ms in total. The remaining portion of the time is attributed to the non-optimized interface used for the communication between the reconstruction-software and the light-sources sequence-control. Figure~\ref{fig6} b) shows the measured height map of a single-shot measurement. The five sub-surfaces are investigated by a region of interest (ROI), containing 200 by 200 pixels  of the 3008~$\times$~3008 pixel camera-sensor, indicated by the white-squares. Alongside, their mean value and standard deviation are stated. The data shows, that each of the logarithmic-spaced sub surfaces is accurately revealed and their resolution of around $\pm0.017$~mm ($1~\sigma$) is close to the typically estimated uncertainty $\Lambda_\mathrm{min}/100\approx0.014$~mm.  
\begin{figure*}[!t]
\centering
\includegraphics[width=182mm]{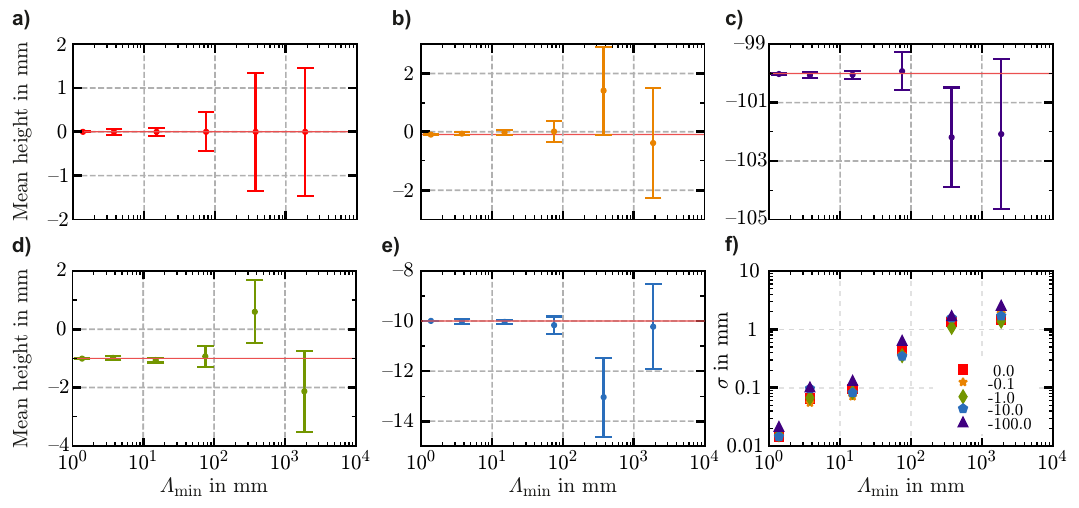}
\caption{Mean height (dots) and standard deviation (error bar) of each sub surface after numerical reconstruction as function of the minimum synthetic wavelength $\Lambda_\mathrm{min}$ used. The solid red lines indicates the nominal height values at 0~mm (a), -0.1~mm (b), -1~mm (c), -10~mm (d) and -100~mm (e). Their standard deviations also plotted in sub-figure (f). }
\label{fig7}
\end{figure*}

Now, we want to provide a more detailed perspective on the cascading process. For this purpose, we start with reconstructing the height map with $\Lambda=1874$~mm only. Then, we add the next shorter synthetic wavelength, i.e. $\Lambda \approx375$~mm, and perform the reconstruction again. By adding more and more smaller synthetic wavelengths, we step-by-step decrease the minimum synthetic wavelength $\Lambda_\mathrm{min}$ to its final value at 1.36~mm for the reconstruction process. For each of these reconstructions, we determine the height values of the 5 sub-surfaces using the same ROI's as indicated in Fig.~\ref{fig6}. Figure~\ref{fig7} shows the mean values along with the respective uncertainties as a function of the minimum synthetic wavelength. One can clearly see that the deviation from the nominal value as well as the uncertainty decrease with the minimum synthetic wavelength. 

It is worth to mention that, it is not strictly necessary to include a measurement at $\Lambda\approx1874$~mm for an unambiguous determination of a height map for this particular sample. The largest height difference is 100~mm. Thus, in principle, the largest synthetic wavelength has to exceed 200~mm. This is already fulfilled with $\Lambda\approx375$~mm. However, we have included the even larger synthetic wavelength in order to demonstrate that the meter-scale synthetic wavelengths generated by the light source can be reliably used for multi-wavelength interferometry.    
\begin{figure}[!b]
\centering
\includegraphics[width=90mm]{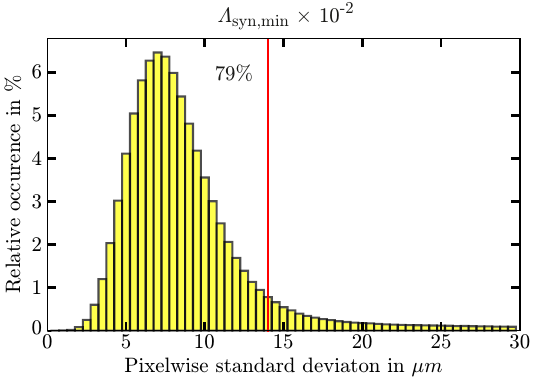}
\caption{Histogram of the pixelwise standard deviation over 10 measurements. 79\% of the pixels standard deviation is below the threshold value of the one hundreth of the finest synthetic wavelength, $\Lambda_{syn,min}$ used, indicated by the red line.}
\label{fig8}
\end{figure}

Finally, we investigate the repeatability of the height determination using the frequency-comb based light source. For this purpose, we conduct 10 consecutive measurements during a total measurement time below 15~s. Figure~\ref{fig8} shows a histogram of the pixelwise standard deviation of the 10 resulting height maps. The graph indicates, that most of the pixels have a standard deviation around 0.007~mm. In the same plot, the red line indicates the typically estimated uncertainty of $\Lambda_\mathrm{min}/100$. For 79~\% of the pixels, the uncertainties are below this value. The vast majority of the outliers are originate from the clamping hole of the sample (marked in white in Fig. \ref{fig6}), from where the reflected signal is too low. The certainty enhancement over $R=10$ measurements indicates, that our system is shot-noise limited, as the standard deviation improves by a factor of $\approx3$ in relation to the single shot measurement. This is expected, because the uncertainty should scale with $1/\sqrt{R}$.   

\section{Conclusion}
\noindent In this work, we proposed and implemented a synthetic wavelength generator, based on a simple but robust frequency comb. Thanks to the ability to choose the sideband order as well as flexibly adjust the modulation frequency, we are able to generate synthetic wavelengths between 1.36~mm and 1874~mm, where the measuring range can be adjusted to different samples. Despite using only two sideband-orders, the synthetic wavelength generator exhibits a dynamic range that exceeds $10^3$. At the same time, a total data acquisition time of 1.5~s for a single shot measurement using 6 cascades is achieved. By maintaining  switching times shorter than 25~ms per synthetic wavelength, the total switching time sums up to 150~ms. In more detail, the examination of the surface of the machined part of aluminum showed, that all height-features between 0.1 and 100~mm are revealed correctly with standard deviations ($1\sigma)$ in the range of 20~\textmu m for a single shot measurement. Averaging over 10 consecutive measurements can improve the uncertainty to the 10~\textmu m level. Thus, our light source allows for measuring meter-sized deformations with an uncertainty of the order of 10~\textmu m which corresponds to a total measurement dynamic range of $10^5$. 

Now, we want to put our results into perspective. In order to obtain a similar dynamic range using a frequency comb combined with a tunable filter ( Fig.~\ref{fig1}b), one would need a frequency comb with 100~MHz spacing and a tunable bandpassfilter with a passband below this spacing, with at least 40~dB out-of-band suppression. Even these two criteria are difficult to achieve at the same time, not to mention challenging practical aspects like repeatability of the passband-tuning.

In relation to the single-sideband modulator architecture ( Fig.~\ref{fig1}a), the presented scheme alleviates the requirements of the maximum rf-frequency by a factor of two, as we use up to the second sideband order and at the same time, the presented approach omits biasing techniques in both, the rf- and the optical domain, which are mandatory for the SSBM-approach.

As outlined in the previous sections, the capabilities of the demonstrated light-source can be enhanced in several aspects, namely power output and switching time. In combination with higher sideband orders available by increasing the rf- driving power,  smaller synthetic wavelengths will become usable and thus the height resolution could reach the single-digit micrometer, resulting in a dynamic range of $10^6$ for the height measurement, enabled by the generation of synthetic wavelengths spanning over a dynamic range of $10^4$ all within one second.
Notably, the light source is composed of field-proven fiber-coupled components, which are all commercially available and widely employed due to their robustness. In combination with the absence of any complicated control-loops, we strongly believe that the proposed architecture has the potential to become an industry standard for fast and reliable interferometric surface determination without additional wavelength-monitoring equipment. 

\section{Disclosures}
\noindent Authors A.S., T.S., J.K. and D.C are employed by the
Fraunhofer Institute for Physical Measurement Techniques,
IPM, which commercially supplies the interferometric sensor
utilized in this work. 
\section{Acknowledgments}
\noindent This work was supported by the German Federal Ministry of
Education and Research, Research Program Quantum Systems
(Grant No. 13N16774). 
\bibliographystyle{IEEEtran}
\bibliography{Ref}

\begin{IEEEbiographynophoto}{Jane Doe}
Biography text here without a photo.
\end{IEEEbiographynophoto}


\end{document}